\documentclass[aps,pra,twocolumn,showpacs,superscriptaddress]{revtex4}
\usepackage{graphicx,color}
\usepackage{amsmath}
\usepackage{amssymb}
\usepackage{bm}

\newcommand{\ket}[1]{|{#1}\rangle}

\newcommand{\bras}[2]{{}_{#2}\hspace*{-0.2mm}\langle{#1}|}

\newcommand{\bracket}[2]{\langle{#1}|{#2}\rangle}

\begin{document}

\title{Resonant Scattering Can Enhance the Degree of Entanglement}


\author{Kazuya Yuasa}
\email[]{kazuya.yuasa@aist.go.jp}
\affiliation{Research Center for Information Security, National Institute of Advanced Industrial Science and Technology (AIST), 1-18-13 Sotokanda, Chiyoda-ku, Tokyo 101-0021, Japan}

\author{Hiromichi Nakazato}
\email[]{hiromici@waseda.jp}
\affiliation{Department of Physics, Waseda University, Tokyo 169-8555, Japan}



\date[]{September 28, 2006}

\begin{abstract}
Generation of entanglement between two qubits by scattering an entanglement mediator is discussed.
The mediator bounces between the two qubits and exhibits a resonant scattering.
It is clarified how the degree of the entanglement is enhanced by the constructive interference of such bouncing processes.
Maximally entangled states are available via adjusting the incident momentum of the mediator or the distance between the two qubits, but their fine tunings are not necessarily required to gain highly entangled states and a robust generation of entanglement is possible.
\end{abstract}


\maketitle

\section{Introduction}
Entanglement plays an essential role in the ideas of quantum information, like quantum computation, quantum communication, quantum cryptography, and so on 
\cite{ref:QuantInfoTextbooks}.
Efficient generation of entanglement hence constitutes an essential element for the realization of such ideas, and various schemes have been proposed.
Entanglement would be most simply and naturally generated by a direct interaction between two entities carrying quantum information 
\cite{ref:Qdots}, 
i.e.\ between two qubits.
There are, however, several setups in which the two qubits are separated from each other, beyond the range of the direct interaction, but an entanglement is to be shared between them. 
In such a case, a ``mediator'' would be convenient to entangle the separated qubits, and several schemes have been investigated theoretically 
\cite{ref:IntMed,ref:cavity,ref:successive0,ref:qpfes,ref:qpfeq,ref:RosannaWJJ,ref:EntScatCostaPRL,ref:EntScatPalma}
and experimentally 
\cite{ref:Haroche,ref:Kuzmich}.

Among these proposals, we here concentrate ourselves on a scheme based on the ``successive'' interactions of a mediator with two qubits to be entangled 
\cite{ref:successive0,ref:qpfes,ref:qpfeq,ref:RosannaWJJ,ref:EntScatCostaPRL,ref:EntScatPalma,ref:Haroche,ref:Kuzmich}.
Mediator X prepared in a specific state is sent to interact successively one by one with qubits A and B, each of which is  prepared in an appropriate initial state, and then the state of X is measured.
If X is found in a specific state after the interactions, we end up with an entanglement between A and B\@.

It has been somewhat standard in such schemes to assume that mediator X interacts with qubit A(B) for a certain time duration $\tau_\text{A(B)}$ 
\cite{ref:successive0,ref:qpfes,ref:qpfeq,ref:RosannaWJJ,ref:EntScatCostaPRL,ref:Haroche},
and the generation of a highly entangled state with a high probability is accomplished by tuning these times, given the coupling constants. 
It is, however, a subtle problem whether the notion of the ``interaction time'' makes sense or not.
It might be valid when the size of a wave packet is small enough compared with the interaction region, but it might not be otherwise.
Even in the former case, no rigorous proof has been given explicitly so far, to the best of the present authors' knowledge.

A rigorous approach to this issue would be to formulate it as a \textit{scattering problem} and such attempts have recently appeared \cite{ref:EntScatCostaPRL,ref:EntScatPalma}.
Qubits A and B are shown to be entangled after \textit{scattering} mediator X\@.
In this approach, the notion of the interaction time is not necessary; actually, such a time is either absent in the stationary-state treatments or automatically given by the physical situation under consideration in the wave-packet scatterings \cite{ref:WavePacketDeltaPot}.
At this point, it is worth noting that X can be reflected by each of the qubits with a certain probability.
This effect is not taken into account in the standard formulations with the interaction times.
The wave reflected by B is partially directed to A and some of its portion is reflected back to B\@.
Such a sequence of bounces between A and B gives rise to a resonant scattering of mediator X\@.

The purpose of the present work is to clarify the effects of the 1D resonant scattering on the entanglement generation \cite{note:NoReso}: resonance can enhance the degree of entanglement.
It is revealed that the distance $d$ between the two qubits is an important parameter to gain an entanglement efficiently, and maximally entangled states are available with finite probabilities by adjusting $d$ or the incident momentum $k$ of X, in a wide parameter region of the coupling constants.
Fine tunings of $k$ or $d$, however, are not necessarily required and robust generation of highly entangled states is possible.

\section{Scattering a Monochromatic Wave}
\begin{figure}[t]
\includegraphics[width=0.4\textwidth]{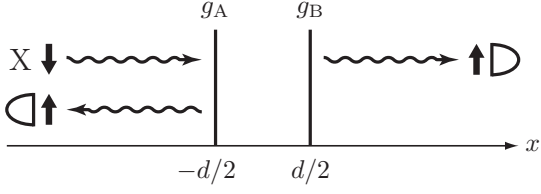}
\caption{Mediator X prepared in $\ket{\downarrow}_\text{X}$ is scattered by qubits A and B placed at $x=-d/2$ and $d/2$, respectively, and initialized in $\ket{\uparrow\uparrow}_\text{AB}$, and $\ket{\uparrow}_\text{X}$ is to be detected by either of the two detectors on both sides, with a certain probability.}
\label{fig:EntGenScat}
\end{figure}

We consider the following Hamiltonian in 1D space:
\begin{align}
H=\frac{p^2}{2m}
&+g_\text{A}(\sigma_+^\text{(X)}\sigma_-^\text{(A)}
+\sigma_-^\text{(X)}\sigma_+^\text{(A)})\delta(x+d/2)\nonumber\\
&+g_\text{B}(\sigma_+^\text{(X)}\sigma_-^\text{(B)}
+\sigma_-^\text{(X)}\sigma_+^\text{(B)})\delta(x-d/2).
\label{eqn:Hamiltonian}
\end{align}
Qubits A and B are placed at $x=-d/2$ and $d/2$, respectively, and the quantum information is encoded in their spin states $\ket{\uparrow}_\text{A(B)}$ and $\ket{\downarrow}_\text{A(B)}$, which are flipped by the ladder operators $\sigma_\mp^\text{(A(B))}$.
A and B are initially prepared in $\ket{\uparrow\uparrow}_\text{AB}$ and mediator X polarized in $\ket{\downarrow}_\text{X}$ is injected from the left.
X, whose position and momentum are represented by the operators $x$ and $p$, respectively, propagates according to the Hamiltonian (\ref{eqn:Hamiltonian}) and is scattered by A and B (Fig.\ \ref{fig:EntGenScat}).
Since the interactions between mediator X and qubits A, B preserve the number of spins in the up state $\ket{\uparrow}$ among A, B, and X, if X after the scattering is detected by either of the two detectors on both sides and found in $\ket{\uparrow}_\text{X}$, either state of A or B is flipped down and an entangled state is generated [see (\ref{eqn:EntStat}) below].
Note that we are looking at an extreme case where the potential barriers are much thinner than the extension of the wave packet of X\@.

Let us find a scattering state of the Hamiltonian (\ref{eqn:Hamiltonian}),
\begin{equation}
H\ket{E}=E\ket{E},
\label{eqn:EigenEq}
\end{equation}
which would describe the scattering of a well-mono\-chromatized incident particle X\@.
We introduce the wave functions
\begin{equation}
u_{{\downarrow};{\uparrow\uparrow}}(x)
=\Bigl(\bras{x{\downarrow}}{\text{X}}\otimes\bras{{\uparrow}{\uparrow}}{\text{AB}}\Bigr)\,\ket{E}
\equiv\bracket{x{\downarrow};{\uparrow}{\uparrow}}{E},\ \text{etc.}
\end{equation}
The solution is given in the form
\begin{equation}
u(x)=\begin{cases}
\medskip
Ie^{ik(x+d/2)}+Re^{-ik(x+d/2)}&(x<-d/2),\\
Te^{ik(x-d/2)}&(x>d/2)
\end{cases}
\end{equation}
for each spin state, where $k$ is the momentum (apart from $\hbar$) of X and $E=\hbar^2k^2/2m$.

We are interested in the entanglement generation from the initialized state $\ket{\uparrow\uparrow}_\text{AB}$ by injecting X in the spin state $\ket{\downarrow}_\text{X}$.
In such a case, only the three wave functions $u_{{\downarrow};{\uparrow\uparrow}}(x)$, $u_{{\uparrow};{\uparrow\downarrow}}(x)$, and $u_{{\uparrow};{\downarrow\uparrow}}(x)$ are involved in the problem.
Imposing the continuity condition on the wave functions, the scattering problem (\ref{eqn:EigenEq}) is solved under the boundary conditions
\begin{equation}
I_{{\downarrow};{\uparrow\uparrow}}=N,\quad
I_{{\uparrow};{\uparrow\downarrow}}=0,\quad
I_{{\uparrow};{\downarrow\uparrow}}=0
\end{equation}
to yield the transmission and reflection amplitudes
\begin{subequations}
\label{eqn:Coefficients}
\begin{align}
T_{{\downarrow};{\uparrow\uparrow}}
&=N\frac{t_\text{A}t_\text{B}e^{ikd}}{1-r_\text{A}r_\text{B}e^{2ikd}},\displaybreak[0]\\
R_{{\downarrow};{\uparrow\uparrow}}
&=N\left(
r_\text{A}
+\frac{t_\text{A}^2r_\text{B}e^{2ikd}}{1-r_\text{A}r_\text{B}e^{2ikd}}
\right),\displaybreak[0]\\
T_{{\uparrow};{\uparrow\downarrow}}
&=R_{{\uparrow};{\uparrow\downarrow}}e^{-ikd}
=N\frac{t_\text{A}f_\text{B}e^{ikd}}{1-r_\text{A}r_\text{B}e^{2ikd}},\displaybreak[0]\\
T_{{\uparrow};{\downarrow\uparrow}}
&=R_{{\uparrow};{\downarrow\uparrow}}e^{ikd}
=N\left(
1+\frac{t_\text{A}r_\text{B}e^{2ikd}}{1-r_\text{A}r_\text{B}e^{2ikd}}
\right)f_\text{A}e^{ikd},
\end{align}
\end{subequations}
where  
\begin{subequations}
\begin{gather}
t_\text{A(B)}=\frac{1}{1+\Omega_\text{A(B)}^2},\qquad
r_\text{A(B)}=-\frac{\Omega_\text{A(B)}^2}{1+\Omega_\text{A(B)}^2},
\displaybreak[0]\\
f_\text{A(B)}=-\frac{i\Omega_\text{A(B)}}{1+\Omega_\text{A(B)}^2},
\end{gather}
\end{subequations}
and $\Omega_\text{A(B)}=mg_\text{A(B)}/\hbar^2k$.
Note that X does not feel A(B) when they are in the same spin states.
The coefficients $t_\text{A(B)}$ and $r_\text{A(B)}$ respectively describe the transmission through and the reflection from A(B) without spin flip and $f_\text{A(B)}$ the transmission/reflection with spin flips, when X is in a different spin state from A(B).

It is interesting to expand the amplitudes in (\ref{eqn:Coefficients}) as power series in $e^{2ikd}$:
\begin{subequations}
\label{eqn:CoefficientsExp}
\begin{align}
T_{{\uparrow};{\uparrow\downarrow}}
&=R_{{\uparrow};{\uparrow\downarrow}}e^{-ikd}\nonumber\\
&=Nt_\text{A}e^{ikd}(
1+r_\text{B}e^{ikd}r_\text{A}e^{ikd}\nonumber\\
&\qquad\qquad\quad\ \ \,%
{}+r_\text{B}e^{ikd}r_\text{A}e^{ikd}r_\text{B}e^{ikd}r_\text{A}e^{ikd}
+\cdots)f_\text{B},\displaybreak[0]\\
T_{{\uparrow};{\downarrow\uparrow}}
&=R_{{\uparrow};{\downarrow\uparrow}}e^{ikd}\nonumber\displaybreak[0]\\
&=N(
1+t_\text{A}e^{ikd}r_\text{B}e^{ikd}\nonumber\\
&\qquad\quad\,
{}+t_\text{A}e^{ikd}r_\text{B}e^{ikd}r_\text{A}e^{ikd}r_\text{B}e^{ikd}
+\cdots)f_\text{A}e^{ikd},
\end{align}
\end{subequations}
etc., where $e^{ikd}$ represents the phase factor gained for the rightward/leftward propagation of X between A and B, and each term of the expansions reveals how X goes back and forth between A and B, before flipping its spin by the interaction with A or B\@.

It is also worth looking at the coefficients $t_\text{A(B)}$, $r_\text{A(B)}$, and $f_\text{A(B)}$ for the single potential A(B) and comparing the present approach with the ordinary formulation with the interaction times \cite{ref:successive0,ref:qpfes,ref:qpfeq,ref:RosannaWJJ,ref:EntScatCostaPRL,ref:Haroche}.
\textit{When we concentrate ourselves on a transmitted particle} (like in the approach with the interaction time), a state of X and A, e.g.\ $\ket{{\downarrow};{\uparrow}}_\text{XA}$, is ``rotated'' after the transmission of X through potential A like
\begin{equation}
\ket{{\downarrow};{\uparrow}}_\text{XA}
\to t_\text{A}\ket{{\downarrow};{\uparrow}}_\text{XA}
+f_\text{A}\ket{{\uparrow};{\downarrow}}_\text{XA},
\end{equation}
apart from the normalization.
The counterpart in the approach with the interaction time reads \cite{ref:qpfeq,ref:RosannaWJJ}
\begin{equation}
\ket{{\downarrow};{\uparrow}}_\text{XA}
\to \cos g_\text{A}\tau_\text{A}
\ket{{\downarrow};{\uparrow}}_\text{XA}
-i\sin g_\text{A}\tau_\text{A}
\ket{{\uparrow};{\downarrow}}_\text{XA},
\end{equation}
and there exists the connection between the two formulations:
\begin{equation}
\cos g_\text{A}\tau_\text{A}
\leftrightarrow\frac{1}{\sqrt{1+\Omega_\text{A}^2}},\quad
\sin g_\text{A}\tau_\text{A}
\leftrightarrow\frac{\Omega_\text{A}}{\sqrt{1+\Omega_\text{A}^2}}.
\label{eqn:correspondence}
\end{equation}
These relations show that $g_\text{A}\tau_\text{A}$ ranges from $0$ to $\pi/2$ and a higher/lower momentum corresponds to a shorter/longer interaction time.

In practice, however, one should also take account of the probability for such event, transmission, to occur.
To achive the complete flip $\sin g_\text{A}\tau_\text{A}=1$, for instance, the correspondence (\ref{eqn:correspondence}) suggests that $\Omega_\text{A}\to\infty$ is required.
But the probability of the transmission
\begin{equation}
|t_\text{A}|^2+|f_\text{A}|^2
=\frac{1}{1+\Omega_\text{A}^2}
\end{equation}
vanishes in this limit, meaning that the complete flip is not possible.
A larger rotation angle $g_\text{A}\tau_\text{A}$ requires a larger $\Omega_\text{A}$ but it is available with a smaller probability.
This restricts the applicability of the formulation with the interaction times, at least for the delta-shaped potential.

\section{Generation of Entanglement}
Now we are ready to discuss the entanglement generation.
If a transmitted/reflected particle in $\ket{\uparrow}_\text{X}$ is detected by the detector on the right/left, an entangled state
\begin{subequations}
\label{eqn:EntStat}
\begin{align}
\ket{\Psi_\text{t}}_\text{AB}
&\propto
T_{\uparrow;\uparrow\downarrow}
\ket{\uparrow\downarrow}_\text{AB}
+T_{\uparrow;\downarrow\uparrow}
\ket{\downarrow\uparrow}_\text{AB}
\label{eqn:EntStatT}
\intertext{(for the former case) or}
\ket{\Psi_\text{r}}_\text{AB}
&\propto
R_{\uparrow;\uparrow\downarrow}
\ket{\uparrow\downarrow}_\text{AB}
+R_{\uparrow;\downarrow\uparrow}
\ket{\downarrow\uparrow}_\text{AB}
\label{eqn:EntStatR}
\end{align}
\end{subequations}
(for the latter) is established.
In the present model, the concurrences \cite{ref:Concurrence} of these states are the same,
\begin{subequations}
\label{eqn:C}
\begin{gather}
C=\frac{2|T_{\uparrow;\uparrow\downarrow}T_{\uparrow;\downarrow\uparrow}|}{|T_{\uparrow;\uparrow\downarrow}|^2+|T_{\uparrow;\downarrow\uparrow}|^2}
=\frac{2|R_{\uparrow;\uparrow\downarrow}R_{\uparrow;\downarrow\uparrow}|}{|R_{\uparrow;\uparrow\downarrow}|^2+|R_{\uparrow;\downarrow\uparrow}|^2}
=\frac{2a}{1+a^2},\displaybreak[0]\\
a
=\left|\frac{T_{\uparrow;\downarrow\uparrow}}{T_{\uparrow;\uparrow\downarrow}}\right|
=\left|\frac{R_{\uparrow;\downarrow\uparrow}}{R_{\uparrow;\uparrow\downarrow}}\right|
=\frac{\Omega_\text{A}}{\Omega_\text{B}}\sqrt{1+4\Omega_\text{B}^2(1+\Omega_\text{B}^2)\sin^2\!kd},
\end{gather}
\end{subequations}
and so are the probabilities of the generations of the entangled states (\ref{eqn:EntStatT}) and (\ref{eqn:EntStatR}), namely the probability of detecting a transmitted particle in the $\ket{\uparrow}_\text{X}$ state and that of detecting a reflected one,
\begin{align}
P&=|T_{\uparrow;\uparrow\downarrow}|^2
+|T_{\uparrow;\downarrow\uparrow}|^2
=|R_{\uparrow;\uparrow\downarrow}|^2
+|R_{\uparrow;\downarrow\uparrow}|^2\nonumber\displaybreak[0]\\
&=\frac{\Omega_\text{A}^2+\Omega_\text{B}^2+4\Omega_\text{A}^2\Omega_\text{B}^2(1+\Omega_\text{B}^2)\sin^2\!kd}%
{(1+\Omega_\text{A}^2+\Omega_\text{B}^2)^2+4\Omega_\text{A}^2\Omega_\text{B}^2(1+\Omega_\text{A}^2)(1+\Omega_\text{B}^2)\sin^2\!kd}.
\label{eqn:P}
\end{align}
Note that the probability of \textit{either} of the two detectors detecting X in the state $\ket{\uparrow}_\text{X}$ is given by $2P$, and therefore the maximal value of $P$ is $1/2$.

\begin{figure}[t]
\includegraphics[height=0.61\textwidth]{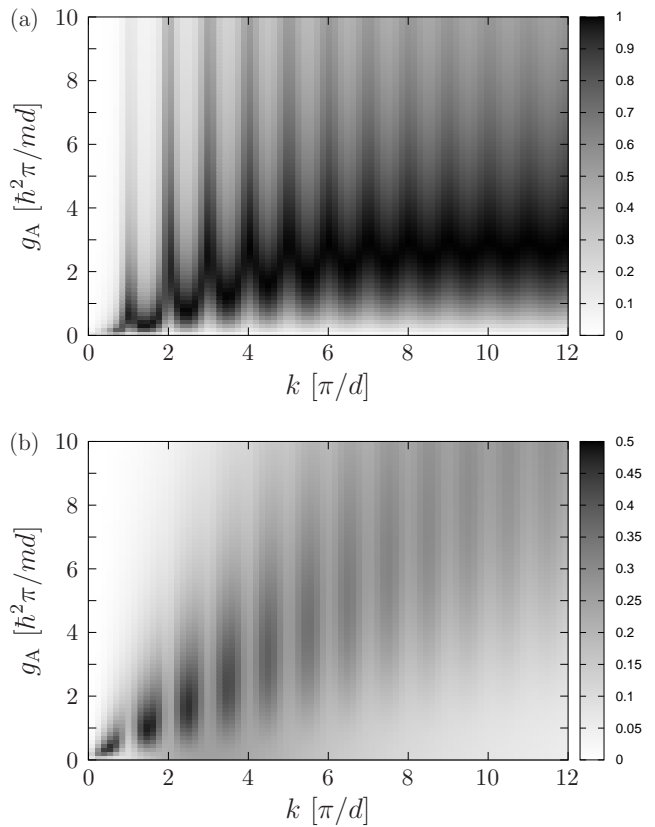}
\caption{(a) The concurrence $C$ given in (\ref{eqn:C}) and (b) the probability $P$ given in (\ref{eqn:P}) as functions of $k$ and $g_\text{A}$ when $g_\text{B}=3\,[\hbar^2\pi/md]$.}
\label{fig:ResoCP}
\end{figure}

\begin{figure}[t]
\includegraphics[width=0.4\textwidth]{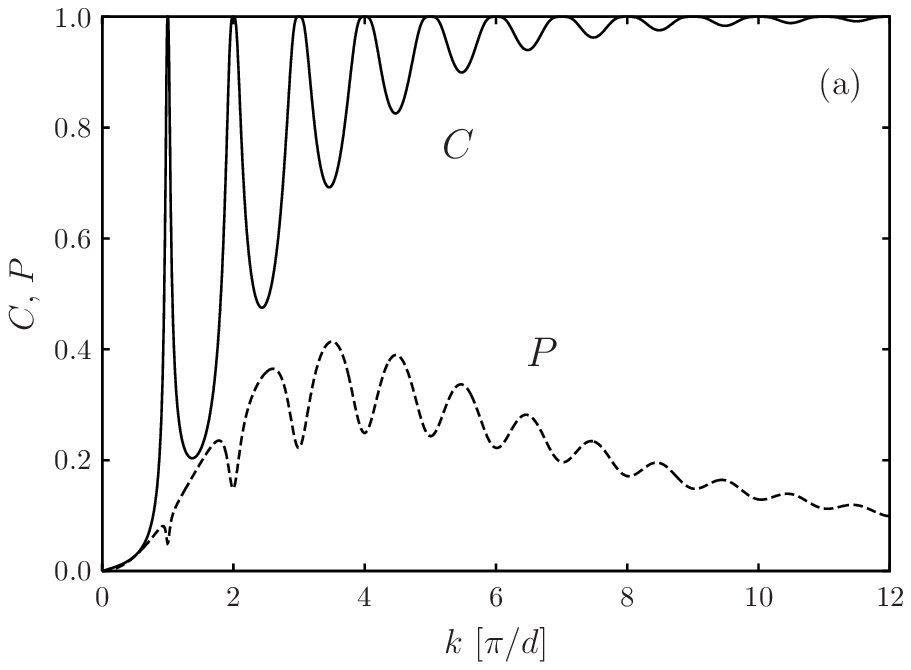}\\
\medskip
\includegraphics[width=0.4\textwidth]{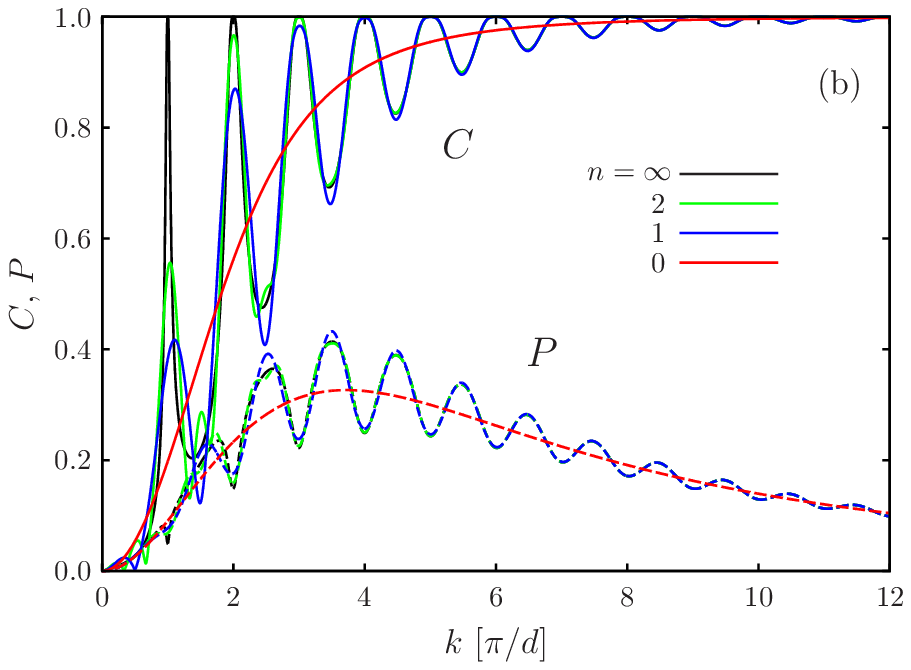}
\caption{(a) The concurrence $C$ given in (\ref{eqn:C}) (solid) and the probability $P$ given in (\ref{eqn:P}) (dashed) as functions of the incident momentum $k$ when $g_\text{A}=g_\text{B}=3\,[\hbar^2\pi/md]$.
(b) The same as (a) but only the contributions up to the $(n+1)$th term in each of the series expansions (\ref{eqn:CoefficientsExp}) are taken into account, where $n$ is the number of bounces between A and B\@.}
\label{fig:ResoCP-gA=gB}
\end{figure}

Notice here that the distance between the two qubits, $d$, enters the formulas through the exponential factor $e^{2ikd}$, which is responsible for the resonant scattering.
See Fig.\ \ref{fig:ResoCP}, where the resonance structures are observed in the momentum dependences of the concurrence $C$ and the probability $P$.
By adjusting the incident momentum $k$, one can generate a highly entangled state with a finite probability.
At the same time, fine tunings of parameters are not necessarily required.
For example, let us look at a case where $g_\text{A}=g_\text{B}$.
Figure \ref{fig:ResoCP-gA=gB}(a) shows that $C$ and $P$ do not oscillate strongly for large $k$ and a high concurrence is available in a wide range of the incident momentum $k$.
This is because the oscillating factor $e^{2ikd}$ represents the bouncing of X between A and B and it always accompanies the reflection coefficients $r_\text{A}$ and $r_\text{B}$, which are reduced for large momentum $k$ and suppress the oscillations.

Let us look more closely at the formulas and see how a highly entangled state is acquired by the resonant tunneling.
Figure \ref{fig:ResoCP-gA=gB}(b) shows that the series expansions in (\ref{eqn:CoefficientsExp}) converge very quickly and the first two contributions, i.e.\ the no- and one-bounce processes, are important.
For $g_\text{A}=g_\text{B}$, the no-bounce process, i.e.\ the first terms in (\ref{eqn:CoefficientsExp}), yields $|T_{\uparrow;\uparrow\downarrow}|<|T_{\uparrow;\downarrow\uparrow}|$ ($|R_{\uparrow;\uparrow\downarrow}|<|R_{\uparrow;\downarrow\uparrow}|$).
If the momentum $k$ is adjusted so as to satisfy $kd=\nu\pi\,(\nu=1,2,\ldots)$, the one-bounce process, i.e.\ the second terms in (\ref{eqn:CoefficientsExp}), interferes with the no-bounce process \textit{constructively} for the $\ket{\uparrow;\uparrow\downarrow}$ component but \textit{destructively} for the $\ket{\uparrow;\downarrow\uparrow}$ component (since $t_\text{A(B)}>0$ while $r_\text{A(B)}<0$).
As a result, the difference between $|T_{\uparrow;\uparrow\downarrow}|$ and $|T_{\uparrow;\downarrow\uparrow}|$ ($|R_{\uparrow;\uparrow\downarrow}|$ and $|R_{\uparrow;\downarrow\uparrow}|$) is reduced by the interference and the concurrence is made close to unity.
In fact, the ratio $a$ introduced in (\ref{eqn:C}) is evaluated up to the one-bounce contribution to be $a=1+\delta a=1+\Omega^6/(1+2\Omega^2+2\Omega^4)$ under the conditions $g_\text{A}=g_\text{B}$ and $kd=\nu\pi\,(\nu=1,2,\ldots)$, and the deviation of the concurrence from unity is about $(\delta a)^2/2$, which is very small for $\Omega\,(=\Omega_\text{A}=\Omega_\text{B})\lesssim1$ (almost perfect compensation).
Other processes with more bounces complete the exact unit concurrence.
Note that, in Fig.\ \ref{fig:ResoCP-gA=gB}, where $g_\text{A}=g_\text{B}=3\,[\hbar^2\pi/md]$, the condition $\Omega\lesssim1$ corresponds to $k\gtrsim3\,[\pi/d]$.
\begin{figure*}[t]
\includegraphics[width=0.43\textwidth]{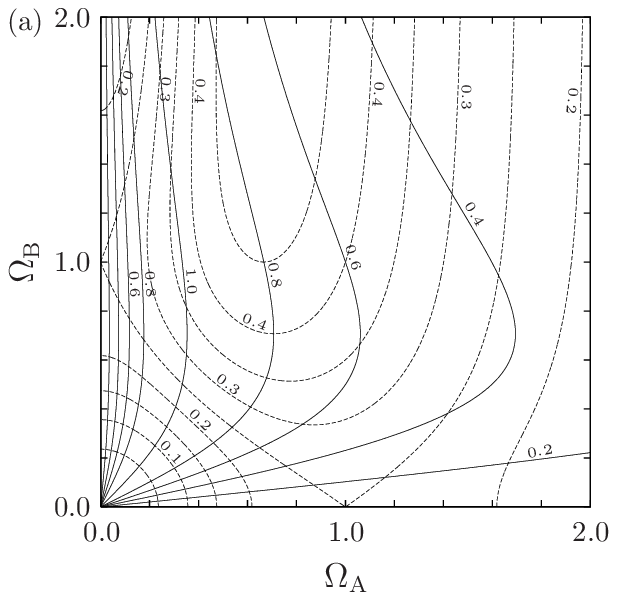}\qquad\qquad
\includegraphics[width=0.43\textwidth]{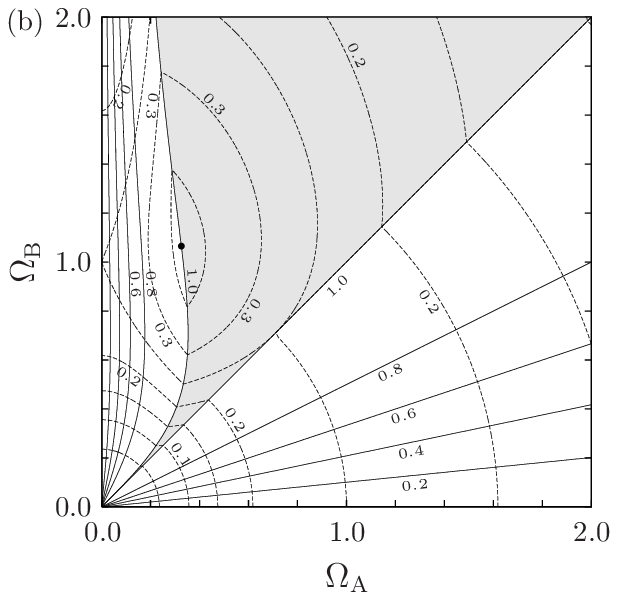}
\caption{(a) Contour plots of the probability $P$ (dashed) given in (\ref{eqn:P}) and the concurrence $C$ (solid) given in (\ref{eqn:C}) as functions of $\Omega_\text{A}$ and $\Omega_\text{B}$ under the resonance condition $\sin^2\!kd=1$. The probability $P$ takes its maximum value $P_\text{max}=1/2$ at $(\Omega_\text{A},\Omega_\text{B})=(1/\sqrt{2},\infty)$, and the concurrence is unity $C=1$ on the line $\Omega_\text{A}=\Omega_\text{B}/(1+2\Omega_\text{B}^2)$.
(b) Contour plots of the optimal concurrence $C$ for given $(\Omega_\text{A},\Omega_\text{B})$ (solid) and the corresponding probability $P$ (dashed) as functions of $\Omega_\text{A}$ and $\Omega_\text{B}$. The unit concurrence $C=1$ is available in the gray region $\Omega_\text{B}/(1+2\Omega_\text{B}^2)\le\Omega_\text{A}\le\Omega_\text{B}$. In the left region $\Omega_\text{A}\le\Omega_\text{B}/(1+2\Omega_\text{B}^2)$, the concurrence $C$ is optimal when $\sin^2\!kd=1$, and in the right region $\Omega_\text{A}\ge\Omega_\text{B}$, it is optimal when $\sin^2\!kd=0$. The optimal probability $P$ for the unit concurrence $C=1$ is realized at the point indicated by a dot, $(\Omega_\text{A},\Omega_\text{B})\simeq(0.33,1.07)$, and is given by $P_\text{opt}\simeq0.37$.}
\label{fig:OptPC}
\end{figure*}

\section{Optimization}
Let us next survey the whole parameter space and discuss how to optimize the concurrence $C$ in (\ref{eqn:C}) and the probability $P$ in (\ref{eqn:P}).
There are essentially three independent (dimensionless) parameters $\Omega_\text{A(B)}=mg_\text{A(B)}/\hbar^2k$ and $kd$.

We first look at the probability (\ref{eqn:P}).
It is possible to show that it is optimal for a given pair $(\Omega_\text{A},\Omega_\text{B})$ when the resonance condition 
\begin{equation}
\sin^2\!kd=1
\label{eqn:Resonance}
\end{equation}
is satisfied.
The probability $P$ under this condition is plotted in Fig.\ \ref{fig:OptPC}(a) as a function of $\Omega_\text{A}$ and $\Omega_\text{B}$, together with the concurrence $C$.
The maximum value of the probability $P$ is given by 
\begin{equation}
P_\text{max}
=\frac{1}{2}\quad\text{at}\quad
(\Omega_\text{A},\Omega_\text{B})=(1/\sqrt{2},\infty),
\end{equation}
while the unit concurrence $C=1$ under the resonance condition (\ref{eqn:Resonance}) is achieved when 
\begin{equation}
\Omega_\text{A}
=\frac{\Omega_\text{B}}{1+2\Omega_\text{B}^2}.
\end{equation}

If one does not stick to a high probability, the concurrence $C$ is optimized as follows.
The concurrence (\ref{eqn:C}) takes its maximum $C=1$ at $a=1$ and decreases monotonically as $a$ leaves this optimal point.
Since 
\begin{equation}
\frac{\Omega_\text{A}}{\Omega_\text{B}}\le a
\le\frac{\Omega_\text{A}}{\Omega_\text{B}}(1+2\Omega_\text{B}^2)
\end{equation}
for a given pair $(\Omega_\text{A},\Omega_\text{B})$, the unit concurrence $C=1$ (i.e.\ $a=1$) requires 
\begin{equation}
\frac{\Omega_\text{A}}{\Omega_\text{B}}\le 1
\le\frac{\Omega_\text{A}}{\Omega_\text{B}}(1+2\Omega_\text{B}^2),\quad
\sin^2\!kd
=\frac{\Omega_\text{B}^2-\Omega_\text{A}^2}%
{4\Omega_\text{A}^2\Omega_\text{B}^2(1+\Omega_\text{B}^2)}.
\end{equation}
See Fig.\ \ref{fig:OptPC}(b), where the region for the unit concurrence $C=1$ is shown, together with the corresponding probability $P$.
In particular, the unit concurrence $C=1$ is always available when $g_\text{A}=g_\text{B}$, by adjusting the incident momentum $k$ or the distance between the two qubits, $d$. 
The optimal probability $P$ for the unit concurrence $C=1$ is realized at the point indicated by a dot in Fig.\ \ref{fig:OptPC}(b), which is evaluated to be
\begin{subequations}
\begin{equation}
P_\text{opt}\simeq0.37
\end{equation}
at 
\begin{gather}
\Omega_\text{B}
=\sqrt{\frac{1+\sqrt[3]{37-3\sqrt{114}}+\sqrt[3]{37+3\sqrt{114}}}{6}}\simeq1.07,\displaybreak[0]\\
\Omega_\text{A}
=\frac{\Omega_\text{B}}{1+2\Omega_\text{B}^2}\simeq0.33,\qquad
\sin^2\!kd=1.
\end{gather}
\end{subequations}

\begin{figure*}
\includegraphics[height=0.61\textwidth]{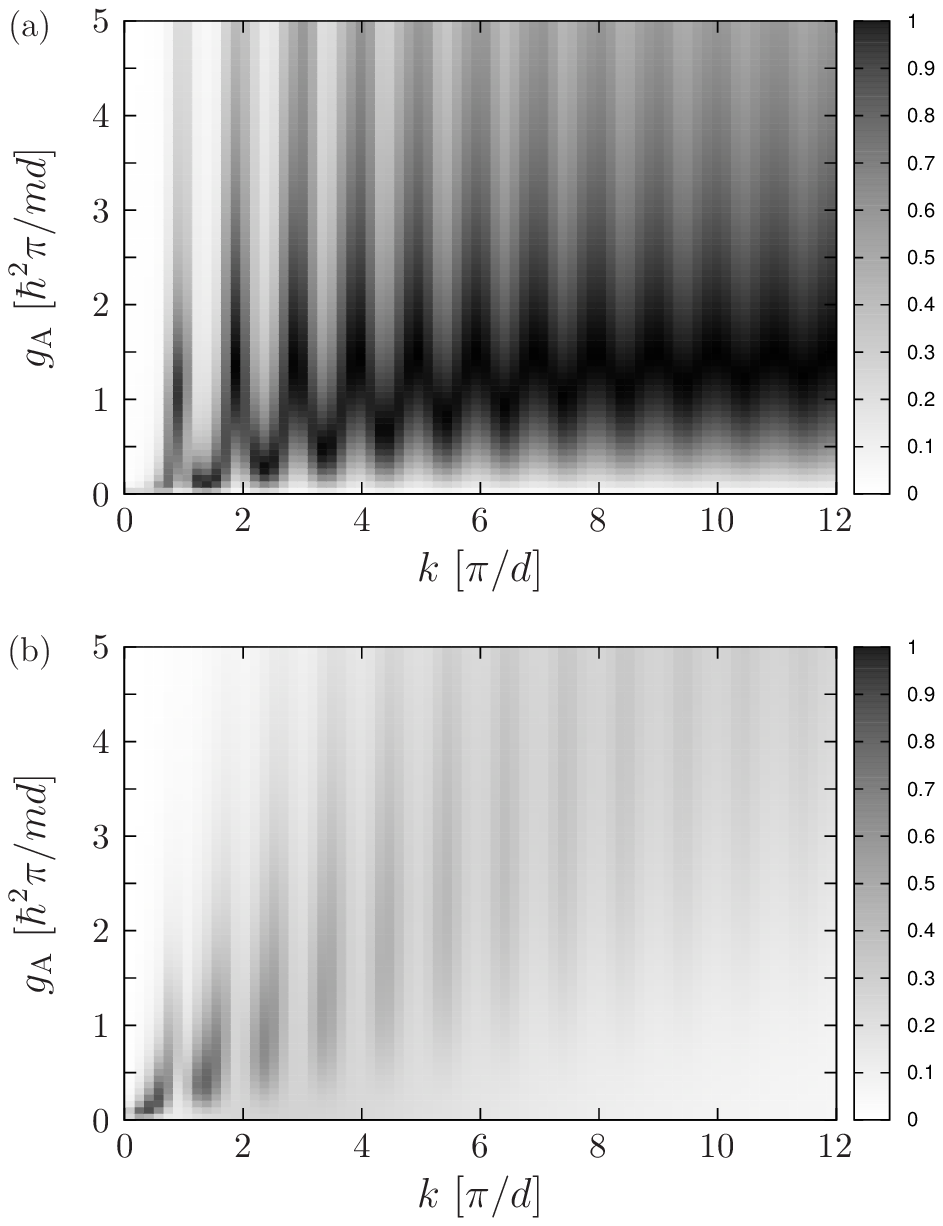}\qquad
\includegraphics[height=0.61\textwidth]{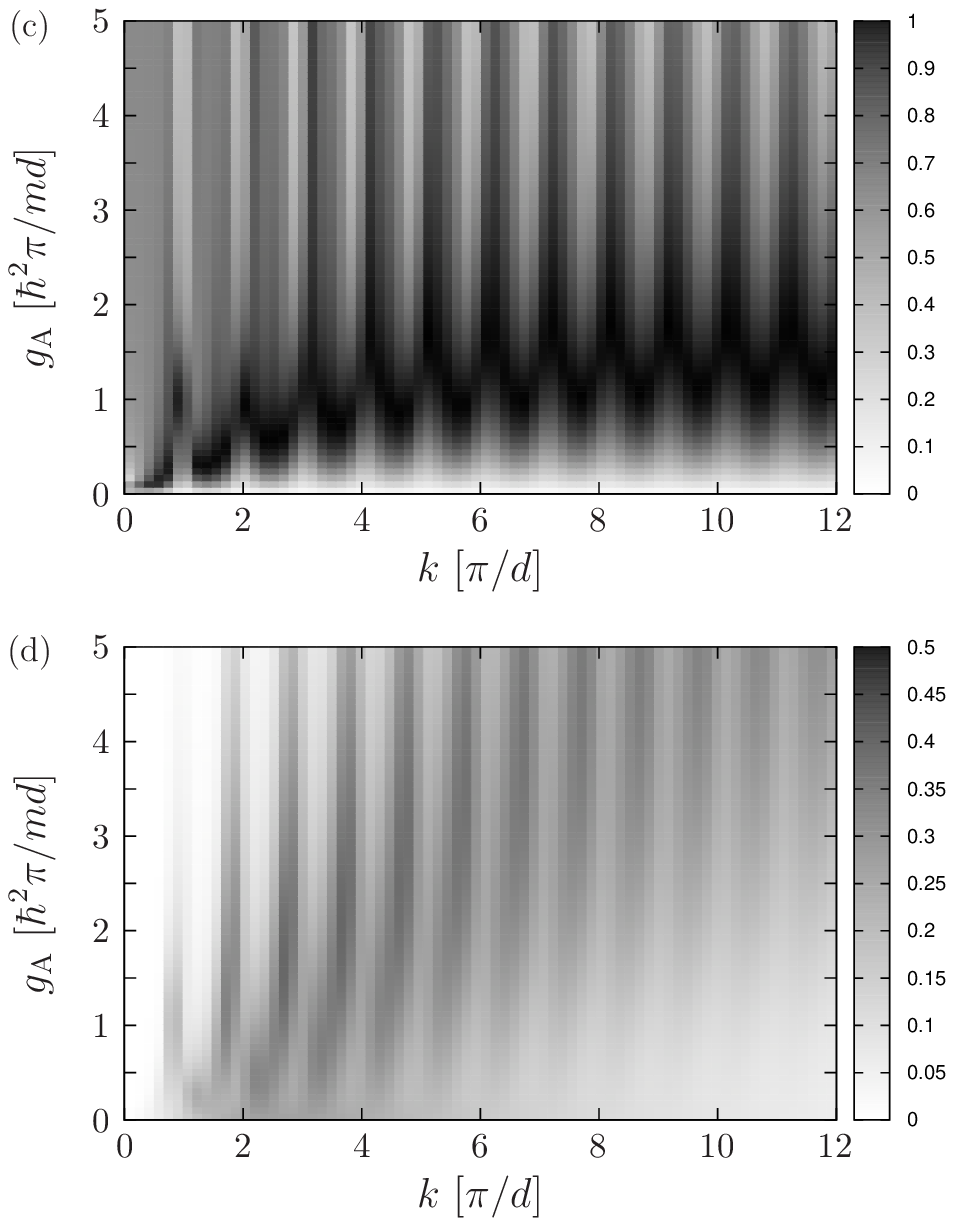}
\caption{The concurrences (above: a,c) and the probabilities (below: b,d) for the Hamiltonian (\ref{eqn:HamiltonianContact}) as functions of $k$ and $g_\text{A}$ with $g_\text{B}=1.5\,[\hbar^2\pi/md]$, when a reflected particle is detected (left: a,b) and when a transmitted particle is detected (right: c,d).}
\label{fig:ResoCP2}
\end{figure*}
\begin{figure*}
\includegraphics[width=0.4\textwidth]{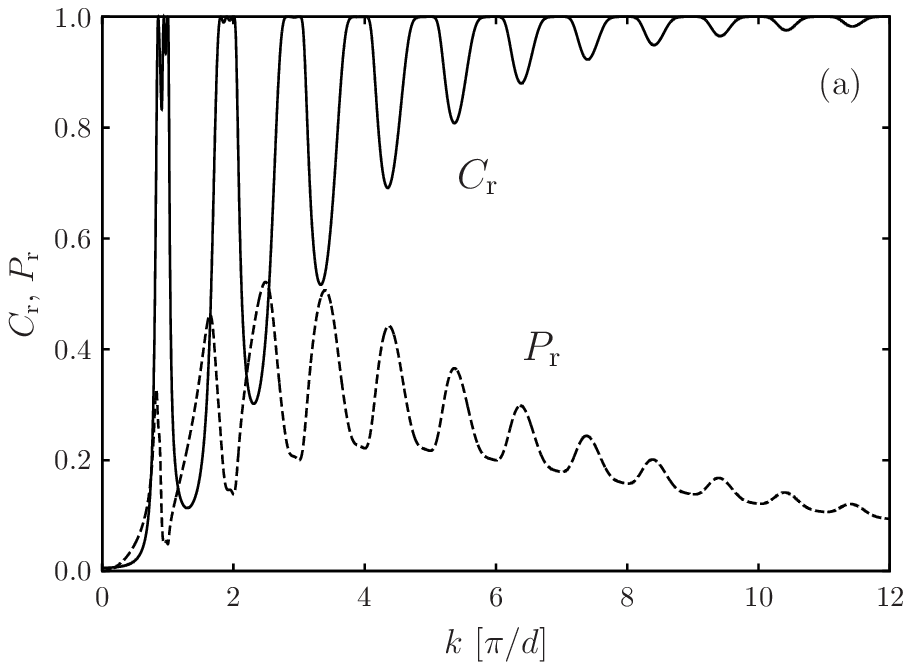}\qquad\qquad\qquad
\includegraphics[width=0.4\textwidth]{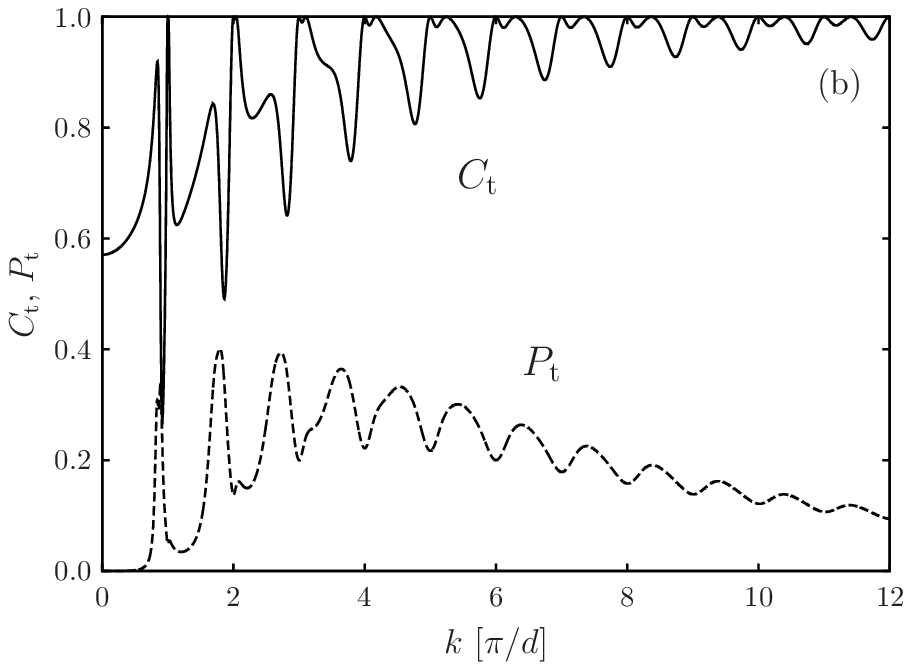}\qquad\mbox{}
\caption{The concurrences (solid) and the probabilities (dashed) for the Hamiltonian (\ref{eqn:HamiltonianContact}) as functions of the incident momentum $k$ with $g_\text{A}=g_\text{B}=1.5\,[\hbar^2\pi/md]$, when a reflected particle is detected (a) and when a transmitted particle is detected (b).}
\label{fig:ResoCP2-gA=gB}
\end{figure*}
\section{Another Model}
\label{sec:Model2}
Similar analyses are possible for the Hamiltonian 
\begin{align}
H=\frac{p^2}{2m}
&+g_\text{A}\bm{\sigma}^\text{(X)}\cdot\bm{\sigma}^\text{(A)}\delta(x+d/2)\nonumber\\
&+g_\text{B}\bm{\sigma}^\text{(X)}\cdot\bm{\sigma}^\text{(B)}\delta(x-d/2).
\label{eqn:HamiltonianContact}
\end{align}
This type of interaction is considered in Ref.\ \cite{ref:EntScatCostaPRL,ref:EntScatPalma}.
In this case, the amplitudes are given by 
\begin{subequations}
\label{eqn:Coefficients2}
\begin{align}
T_{\downarrow;\uparrow\uparrow}
&=N\frac{\tilde{t}_\text{A}\tilde{t}_\text{B}e^{ikd}}{1-\tilde{r}_\text{A}\tilde{r}_\text{B}e^{2ikd}},\displaybreak[0]\\
R_{\downarrow;\uparrow\uparrow}
&=N\left(
\tilde{r}_\text{A}
+\frac{\tilde{t}_\text{A}^2\tilde{r}_\text{B}e^{2ikd}}{1-\tilde{r}_\text{A}\tilde{r}_\text{B}e^{2ikd}}
\right),\\
T_{\uparrow;\uparrow\downarrow}
&=N\frac{\tilde{t}_\text{A}e^{ikd}}{1-\tilde{r}_\text{A}\tilde{r}_\text{B}e^{2ikd}}f_\text{B}\left(
1+\frac{r_\text{A}'t_\text{B}e^{2ikd}}{1-r_\text{A}'r_\text{B}e^{2ikd}}
\right),\\
R_{\uparrow;\uparrow\downarrow}
&=N\frac{\tilde{t}_\text{A}e^{ikd}}{1-\tilde{r}_\text{A}\tilde{r}_\text{B}e^{2ikd}}f_\text{B}\frac{t_\text{A}'e^{ikd}}{1-r_\text{A}'r_\text{B}e^{2ikd}},\displaybreak[0]\\
T_{\uparrow;\downarrow\uparrow}
&=N\left(
1+\frac{\tilde{t}_\text{A}\tilde{r}_\text{B}e^{2ikd}}{1-\tilde{r}_\text{A}\tilde{r}_\text{B}e^{2ikd}}
\right)f_\text{A}\frac{t_\text{B}'e^{ikd}}{1-r_\text{A}r_\text{B}'e^{2ikd}},\displaybreak[0]\\
R_{\uparrow;\downarrow\uparrow}
&=N\left(
1+\frac{\tilde{t}_\text{A}\tilde{r}_\text{B}e^{2ikd}}{1-\tilde{r}_\text{A}\tilde{r}_\text{B}e^{2ikd}}
\right)\nonumber\\
&\qquad\qquad\qquad
{}\times f_\text{A}\left(
1+\frac{t_\text{A}r_\text{B}'e^{2ikd}}{1-r_\text{A}r_\text{B}'e^{2ikd}}
\right),
\end{align}
\end{subequations}
with 
\begin{subequations}
\begin{align}
t_\text{A(B)}
&=\frac{1-i\Omega_\text{A(B)}}{(1+i\Omega_\text{A(B)})(1-3i\Omega_\text{A(B)})},\displaybreak[0]\\
r_\text{A(B)}
&=\frac{i\Omega_\text{A(B)}(1+3i\Omega_\text{A(B)})}{(1+i\Omega_\text{A(B)})(1-3i\Omega_\text{A(B)})},\displaybreak[0]\\
f_\text{A(B)}
&=-\frac{2i\Omega_\text{A(B)}}{(1+i\Omega_\text{A(B)})(1-3i\Omega_\text{A(B)})},\displaybreak[0]\\
t_\text{A(B)}'
&=\frac{1}{1+i\Omega_\text{A(B)}},\quad
r_\text{A(B)}'=-\frac{i\Omega_\text{A(B)}}{1+i\Omega_\text{A(B)}},
\end{align}
\end{subequations}
and
\begin{subequations}
\label{eqn:RenCoefficients2}
\begin{gather}
\tilde{t}_\text{A(B)}
=t_\text{A(B)}
+\Sigma_\text{A(B)},\quad
\tilde{r}_\text{A(B)}
=r_\text{A(B)}+\Sigma_\text{A(B)},\displaybreak[0]\\
\Sigma_\text{A(B)}
=\frac{f_\text{A(B)}^2r_\text{B(A)}'e^{2ikd}}{1-r_\text{A(B)}r_\text{B(A)}'e^{2ikd}}.
\end{gather}
\end{subequations}
In contrast to the previous model, X feels potential A(B) even when X is in the same spin state as A(B)\@.
$t_\text{A(B)}'$ and $r_\text{A(B)}'$ are the coefficients for the transmission through and the reflection from the single potential A(B) without spin flip when X is in the same spin state as A(B), $t_\text{A(B)}$ and $r_\text{A(B)}$ are those when X is in a different spin state from A(B), and $f_\text{A(B)}$ describes the transmission and the reflection with spin flip when X is in a different spin state from A(B).
For this model, the concurrence $C_\text{t}$ and the probability $P_\text{t}$ by detecting a transmitted particle in $\ket{\uparrow}_\text{X}$ on the right are different from $C_\text{r}$ and $P_\text{r}$ by detecting a reflected one on the left.
Again, the denominators in (\ref{eqn:Coefficients2}) and (\ref{eqn:RenCoefficients2}) represent bouncing of X between A and B (which becomes clear by expanding them as power series in $e^{2ikd}$) and give rise to the resonant tunneling.
Similar resonance structures to the previous example are observed in the concurrences and the probabilities, as shown in Figs.\ \ref{fig:ResoCP2} and \ref{fig:ResoCP2-gA=gB}.
The oscillations of the concurrences and the probabilities are reduced for a large momentum $k$ and a robust entanglement generation is available (Fig.\ \ref{fig:ResoCP2-gA=gB}), due to the same mechanism as in the previous example.

\section{Summary}
In this article, we have investigated the entanglement generation by the resonant scattering.
The resonance effects are clarified and the optimization of the entanglement generation is discussed.
The interference of the bouncing and non-bouncing processes can enhance entanglement.
Maximally entangled states are available with finite probabilities in a wide parameter region.
The degree of entanglement is optimized by adjusting the momentum of the mediator, but its fine tuning is not necessarily required.

One of the interesting future subjects would be to clarify the effect of the size of a wave packet on the generation of entanglement.
Scattering of plane waves is discussed in this article, but the models are solvable also for wave packets \cite{ref:WavePacketDeltaPot}.
It is possible to discuss the resonant scattering fully dynamically.
Introduction of the width of the potential would also be interesting to explore the validity of the ordinary formulation with the interaction times.

\section*{Acknowledgements}
This work is partly supported by the bilateral Italian--Japanese Projects II04C1AF4E on ``Quantum Information, Computation and Communication'' of the Italian Ministry of Instruction, University and Research, and 15C1 on ``Quantum Information and Computation'' of the Italian Ministry for Foreign Affairs, by the Grant for The 21st Century COE Program at Waseda University and the Grants-in-Aid for Scientific Research on Priority Areas (No.\ 13135221) and for Young Scientists (B) (No.\ 18740250) from the Ministry of Education, Culture, Sports, Science and Technology, Japan, and by the Grants-in-Aid for Scientific Research (C) (No.\ 18540292) from the Japan Society for the Promotion of Science.


\begin{thebibliography}{10}
\bibitem{note:NoReso}
The resonance effect is not clarified in Ref.\ \cite{ref:EntScatCostaPRL}. In
  fact, the distance between the two scatterers, $d$, is not contained in the
  formulas in Ref.\ \cite{ref:EntScatCostaPRL}, while it is the important
  parameter for the resonance. Reference \cite{ref:EntScatPalma} briefly
  discusses the entanglement generation in the same system as that discussed in Sec.\ \ref{sec:Model2} of 
  this article, but the parameter $kd$ (which is $kx_0$ in Ref.\
  \cite{ref:EntScatPalma}) is fixed at $kd=\nu\pi\,(\nu=1,2,\ldots)$, and the
  effect of the resonant scattering on the entanglement generation is not fully
  clarified.
\end{thebibliography}
\end{document}